\documentclass[superscriptaddress,amsmath,amssymb,aps,prl,twocolumn,floatfix]{revtex4-1}

\usepackage{graphicx}
\usepackage{dcolumn}
\usepackage{bm}
\usepackage{xcolor}
\usepackage{amsmath}
\usepackage{chngcntr}
\usepackage{nicefrac}
\usepackage{hyperref}
\usepackage{natbib}
\usepackage[utf8]{inputenc}
\begin{document}

\title{Interlayer hybridization enables superconductivity in bilayer nickelates}

\author{Shilong Zhang}\thanks{These authors contributed equally to this work.}
\affiliation{International Center for Quantum Materials, School of Physics, Peking University, Beijing 100871, China}

\author{Meng Zhang}\thanks{These authors contributed equally to this work.}
\affiliation{School of Physics, State Key Laboratory for Extreme Photonics and Instrumentation, Zhejiang University, Hangzhou 310027, China}

\author{Qilin Luo}\thanks{These authors contributed equally to this work.}
\affiliation{National Laboratory of Solid State Microstructures and Department of Physics, Nanjing University, Nanjing 210093, China}

\author{Zihao Tao}
\affiliation{International Center for Quantum Materials, School of Physics, Peking University, Beijing 100871, China}

\author{Hsiao-Yu Huang}
\affiliation{National Synchrotron Radiation Research Center, Hsinchu 30076, Taiwan}

\author{Kunhao Li}
\affiliation{International Center for Quantum Materials, School of Physics, Peking University, Beijing 100871, China}

\author{Ganesha Channagowdra}
\affiliation{National Synchrotron Radiation Research Center, Hsinchu 30076, Taiwan}

\author{Jie Li}
\affiliation{National Laboratory of Solid State Microstructures and Department of Physics, Nanjing University, Nanjing 210093, China}

\author{Junchi Fu}
\affiliation{School of Physics, State Key Laboratory for Extreme Photonics and Instrumentation, Zhejiang University, Hangzhou 310027, China}

\author{Di-Jing Huang}
\affiliation{National Synchrotron Radiation Research Center, Hsinchu 30076, Taiwan}

\author{Yanwu Xie}
\email{ywxie@zju.edu.cn}
\affiliation{School of Physics, State Key Laboratory for Extreme Photonics and Instrumentation, Zhejiang University, Hangzhou 310027, China}

\author{Yi Lu}
\email{yilu@nju.edu.cn}
\affiliation{National Laboratory of Solid State Microstructures and Department of Physics, Nanjing University, Nanjing 210093, China}

\author{Yingying Peng}
\email{yingying.peng@pku.edu.cn}
\affiliation{International Center for Quantum Materials, School of Physics, Peking University, Beijing 100871, China}
\affiliation{Collaborative Innovation Center of Quantum Matter, Beijing 100871, China}

\date{\today}

\begin{abstract}

Ruddlesden--Popper nickelates offer a new route to high-temperature superconductivity beyond the cuprates and iron-pnictides. 
However, the electronic reorganization that enables superconductivity in bilayer nickelates remain unresolved, largely due to the difficulty of directly probing the superconducting phase.
Here, we overcome this limitation by stabilizing superconducting (La,Pr)$_3$Ni$_2$O$_7$ thin films with a protective capping layer, thereby enabling direct spectroscopic access via X-ray absorption and resonant inelastic X-ray scattering. We resolve the evolution of in-plane and out-of-plane electronic states, spin and orbital excitations, and spin-density-waves across insulating, superconducting, and metallic regimes. 
Combining experimental results with theoretical analysis, we show that the in-plane $d_{x^2-y^2}$ states form an itinerant backbone, while superconductivity emerges only when coherent $d_{z^2}$--$p_z$--$d_{z^2}$ interlayer hybridization develops, accompanied by suppressed static spin order and strongly damped spin excitations.  
Oxygen stoichiometry and epitaxial strain both act on this interlayer channel, placing superconductivity within a narrow window of interlayer coherence and correlation strength. These findings identify the microscopic ingredients required for superconductivity in bilayer nickelates and provide a multiorbital picture of its emergence.

\end{abstract}

\maketitle

The recent discovery of high-temperature superconductivity in Ruddlesden--Popper (RP) nickelates has opened a new frontier in the study of unconventional superconductivity beyond cuprates. Bilayer nickelates such as La$_3$Ni$_2$O$_7$ exhibit superconductivity with transition temperatures approaching 80--90 K under high pressure~\cite{SC_LNO327_nat2023,SC_LPNO4310_Nat2024,SC_poly_LNO327_PRX2024,SC_LNO5310_1212_NP2025,SC_96K_LSmNO_nat2025_Zhang,SC_LPNO2127_poly_Nat2024}. Through epitaxial strain engineering and chemical substitution, superconductivity has recently been stabilized in thin films at ambient pressure with $T_c$ up to $\sim$60 K~\cite{SC_LNOfilm_Nat2024_Hwang,SC_LPNOfilm_40K_Nat2025_Chen,SCdome_growth_LSNO_NM2025,SC_LPNO1227_stable_AM2025,SC_film_60K_arxiv_QKX}, revealing a rich phase diagram spanning insulating, metallic and superconducting regimes. Unlike cuprates~\cite{LSCO_XAS_doping_PRL1991_CTC,LSCO_XAS_pol_PRL1992_CTC,ZRS_cuprate} or infinite-layer nickelates~\cite{SC_RNO112_nat2019}, these systems host a multiorbital electronic structure involving both $d_{x^2-y^2}$ and $d_{z^2}$ orbitals, coupled through apical oxygen atoms between adjacent NiO$_2$ planes~\cite{SC_LNO327_nat2023,SC_theory_PRL2024_WCJ,LNO327_Theory_pair_PRL2023_YF}. This distinctive configuration has motivated extensive theoretical proposals invoking interlayer magnetic exchange and orbital-selective pairing as key ingredients of superconductivity~\cite{LNO327_Theory_dz_pair_PRB2023_WQH,LNO327_theory_interlayer_PRB2023_ZFC,SC_theory_PRL2024_WCJ,LNO327_Theory_pair_PRL2023_YF,theory_PRB2023_lecherman,theory_PRL2023_werner,theory_PRB2024_yangyifeng,theory_PRL2024_kuroki,theory_NC2024_Zhangyang,theory_PRL2023_yaodaoxin,theory_PRL2024_SG,theory_PRL2024_WZY,theory_PRB2024_LZY,theory_npj2024_YDX}. Nevertheless, the microscopic origin of superconductivity remains intensely debated, with key open questions concerning the roles of structure~\cite{SC_LNO327_nat2023,SC_LPNO2127_poly_Nat2024,SC_higherP_LNO327_NSR_WM2025,SC_LNOfilm_Nat2024_Hwang,SC_LPNOfilm_40K_Nat2025_Chen,RT_FL_NC2026_hwang,SC_LPNO2127_film_NM_Hwang,SCdome_growth_LSNO_NM2025,SC_gap_ARPES_gamma_arxiv_Xue,SC_LAO_film_AM_Hwang,NV_LNO327_Nat2026,struct_EELS_arxiv2025}, oxygen stoichiometry ~\cite{SC_LNOfilm_Nat2024_Hwang,SC_LPNOfilm_40K_Nat2025_Chen,O_content_crystal,O_overdop_JSSC2026,O_stabiliz_arxiv2025,Oxygen_EELS_LNO327_Nat2024,oxygen_EELS_LPNO2127_NM2025} and competing density-wave orders~\cite{SDW_muSR_LNO327_PRL_2024,SDW_pressure_muSR_LNO327_NP2025,SDW_pressure_muSR_LNO327_PRR2025,SDW_RIXS_LNO327_NC_2024,SDW_tetra_LNO327_NC2025}.

A major obstacle is the lack of direct spectroscopic access to the superconducting state. In bulk crystals, superconductivity emerges only under high pressure~\cite{SC_LNO327_nat2023,SC_LPNO4310_Nat2024,SC_poly_LNO327_PRX2024,SC_LNO5310_1212_NP2025,SC_96K_LSmNO_nat2025_Zhang,SC_LPNO2127_poly_Nat2024}, restricting the applicability of most spectroscopy probes. In thin films, although superconductivity can be realized at ambient pressure, it is highly sensitive to oxygen loss and structural inhomogeneity, often leading to phase separation and degraded superconducting properties~\cite{SC_LNOfilm_Nat2024_Hwang,SC_LPNOfilm_40K_Nat2025_Chen,SC_LPNO1227_stable_AM2025,SC_LPNO2127_film_NM_Hwang}. As a result, the intrinsic electronic and magnetic structures in the RP-phase superconductors remain largely unresolved.

Here, by optimizing growth conditions and employing a protective capping layer, we realize robust superconductivity in LaPr$_2$Ni$_2$O$_7$ thin films that persists throughout X-ray absorption spectroscopy (XAS) and resonant inelastic X-ray scattering (RIXS) measurements, enabling direct spectroscopic access to the insulating, superconducting, and metallic regimes. We track the evolution of orbital-selective electronic states, spin and orbital excitations, and spin-density-wave correlations across this phase diagram, revealing the electronic reconstruction from a localized to an itinerant regime. The superconducting films are characterized by enhanced ligand-hole character, a reorganization of out-of-plane $d_{z^2}$ states, suppression of static spin order, and the emergence of strongly damped, itinerant spin excitations. Combining experiment with theory, we show that oxygen stoichiometry and epitaxial strain cooperatively tune the interlayer coupling and hole density, thereby controlling $d_{z^2}\!-\!p_z\!-\!d_{z^2}$ hybridization and weakening competing orders. Together, these results provide a coherent spectroscopic framework for the insulator--superconductor--metal evolution in bilayer nickelates and highlight the central role of interlayer hybridization in enabling superconductivity.

\begin{figure}[htbp]
    \centering
    \includegraphics [width = \columnwidth]
    {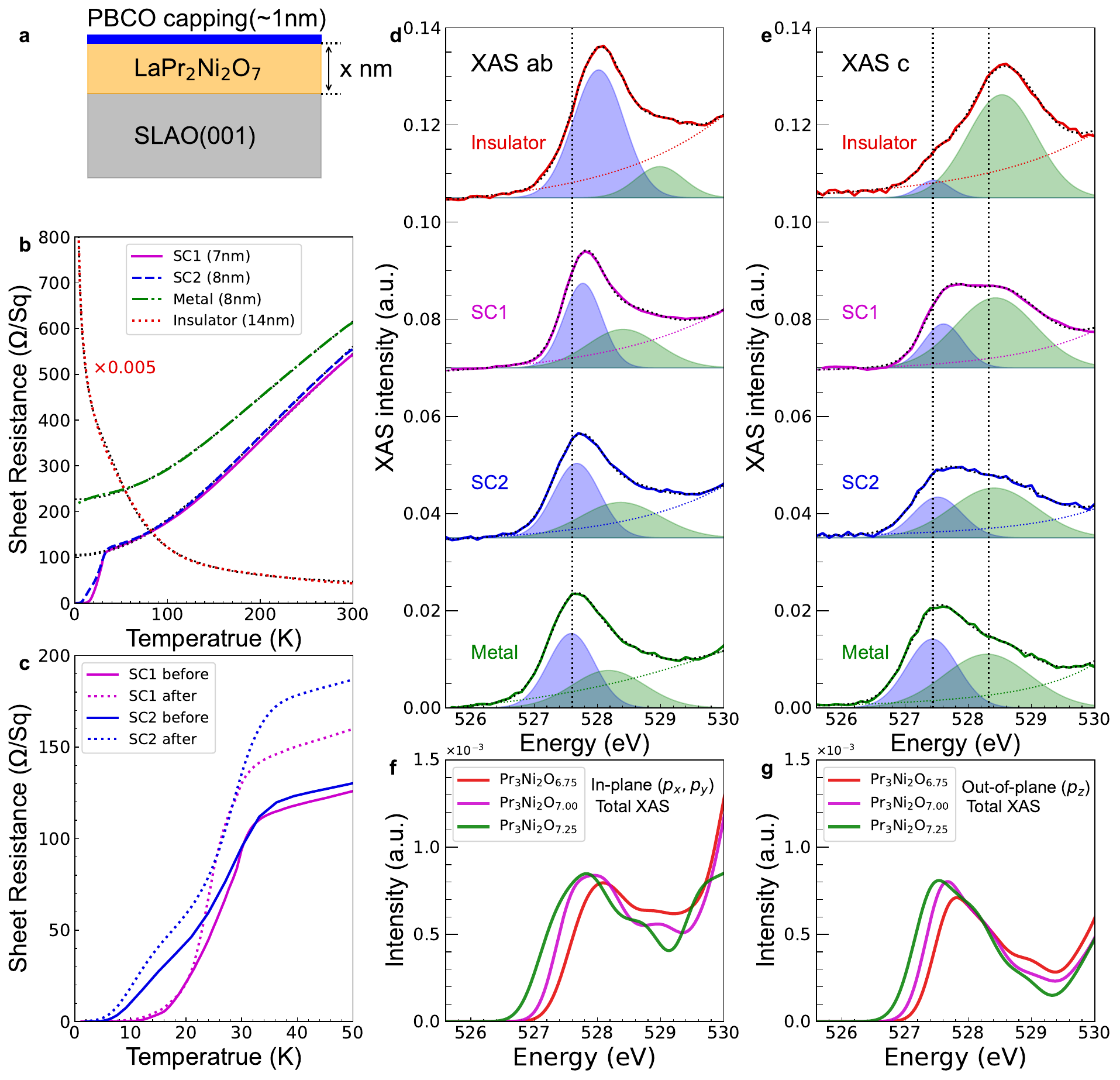}
    \caption{\textbf{Resistance and O $K$-edge XAS of LaPr$_2$Ni$_2$O$_7$ thin films.}
\textbf{a}, Schematic of the thin-film structure, including an $\sim$1 nm amorphous PrBa$_2$Cu$_3$O$_7$ (PBCO) capping layer for surface protection. 
\textbf{b}, Resistance--temperature ($R$--$T$) curves for two superconducting samples, a metallic sample, and an insulating sample; film thicknesses are indicated in the legend. 
\textbf{c}, $R$--$T$ curves of two superconducting samples before and after XAS and RIXS measurements, demonstrating the robustness of superconductivity. 
\textbf{d}, In-plane O $K$-edge XAS pre-edge spectra. The dashed line shows the fitted background, and the shaded curves represent Gaussian components used to fit the pre-edge features. From the insulating to superconducting to metallic samples, the peak positions shift to lower energy, consistent with increasing hole doping.
\textbf{e}, Out-of-plane O $K$-edge XAS pre-edge spectra and corresponding fits. 
\textbf{f,g}, Calculated in-plane and out-of-plane O $K$-edge XAS spectra for different oxygen configurations.
}
    \label{fig1}
\end{figure}

~\\
\leftline{\large{\textbf{Results}}}
\leftline{\textbf{Sample design and transport characterization}}
LaPr$_2$Ni$_2$O$_7$ thin films were grown on SLAO(001) substrates and capped with a 1~nm amorphous PrBa$_2$Cu$_3$O$_7$ (PBCO) layer to preserve oxygen contents (Fig.~1a). By tuning growth conditions and thickness, we obtained samples spanning insulating, superconducting and metallic regimes. The insulating sample is 14~nm thick, whereas the metallic and superconducting samples are 8~nm and 7--8~nm thick, respectively. As shown in Fig.~1b, the two superconducting samples (SC1 and SC2) exhibit identical onset transition temperatures ($T_{c,\mathrm{onset}}$) at $\sim$30 K, while SC1 shows a higher zero-resistance temperature ($T_{c,\mathrm{zero}}$) at $\sim$10 K, indicating improved phase coherence. The normal-state transport further distinguishes these regimes: the metallic and superconducting samples above $T_c$ are well described by a parallel-resistor model $1/R = 1/(R_0 + AT^2) + 1/R_{\mathrm{max}}$~\cite{RT_FL_NC2026_hwang}, consistent with Fermi-liquid-like behavior. In contrast, the insulating sample follows $1/R = 1/(R_{\mathrm{local}} e^{(T_{\mathrm{local}}/T)^{1/4}}) + 1/(R_{\mathrm{gap}} e^{T_{\mathrm{gap}}/T})$, where first term corresponds to three-dimensional Mott variable-range hopping of disorder-localized carriers, while the second term captures thermally activated transport across a density-wave gap~\cite{RT_gap_RMP1998}. To verify the robustness of superconductivity during spectroscopy, we compared the resistance of SC1 and SC2 before and after XAS and RIXS measurements (Fig.~1c). Although the overall resistance increases due to partial degradation, both $T_{c,\mathrm{onset}}$ and $T_{c,\mathrm{zero}}$ remain unchanged, demonstrating that superconductivity is preserved throughout the measurements.

~\\
\leftline{\textbf{Polarization-resolved O K-edge XAS}}
To elucidate the role of oxygen distribution and stoichiometry in shaping the electronic structure, we resolve the O K-edge XAS into in-plane and out-of-plane components. The in-plane response ($E \parallel ab$), arising from O $p_{x,y}$ orbitals hybridized with Ni $d_{x^2-y^2}$ states, is shown in Fig.~1d. From the insulating to superconducting and metallic samples, the spectral peak shift systematically to lower energy. Assuming an approximately constant O 1$s$ core level, this trend indicates a downward shift of unoccupied states, consistent with progressive hole doping. The out-of-plane component ($E \parallel c$), associated with O $p_z$ orbitals hybridized with Ni $d_{z^2}$ states, reveals a richer structure (Fig.~1e). 
The spectrum comprises two components separated by $\sim$1~eV, both of which shift to lower energy with increasing doping, while spectral weight transfers from the higher- to the lower-energy feature. This distinct evolution indicates substantial reconstruction of the $d_{z^2}$-derived states and highlights the central role of out-of-plane orbitals in the electronic reconstruction across the phase diagram.

To gain microscopic insight, we performed DFT calculations of the O K-edge XAS for varying oxygen configurations, including inner-apical vacancies and interstitial oxygens (Figs.~1f,g). The calculations reproduce the overall energy shift, confirming its origin in hole doping. The in-plane spectra are captured reasonably well, consistent with the more itinerant character of Ni $d_{x^2-y^2}$ states. 
By contrast, DFT only captures the lower-energy out-of-plane component derived dominantly from the inner-apical O $p_z$ states, while the higher-energy feature is absent. The missing high-energy spectral weight points to correlation effects beyond a single-particle description. Given the strong hybridization between O $p_z$ and the more localized Ni $d_{z^2}$ orbitals~\cite{theory_PRB2023_lecherman,theory_PRL2023_werner,theory_PRB2024_yangyifeng}, this additional spectral weight likely reflects incoherent excitations in the $d_{z^2}$ channel at an energy scale comparable to the charge-transfer gap~\cite{SDW_RIXS_LNO327_NC_2024,PhysRevB_111_014515}.

\begin{figure}[htbp]
    \centering
    \includegraphics [width = \columnwidth]
    {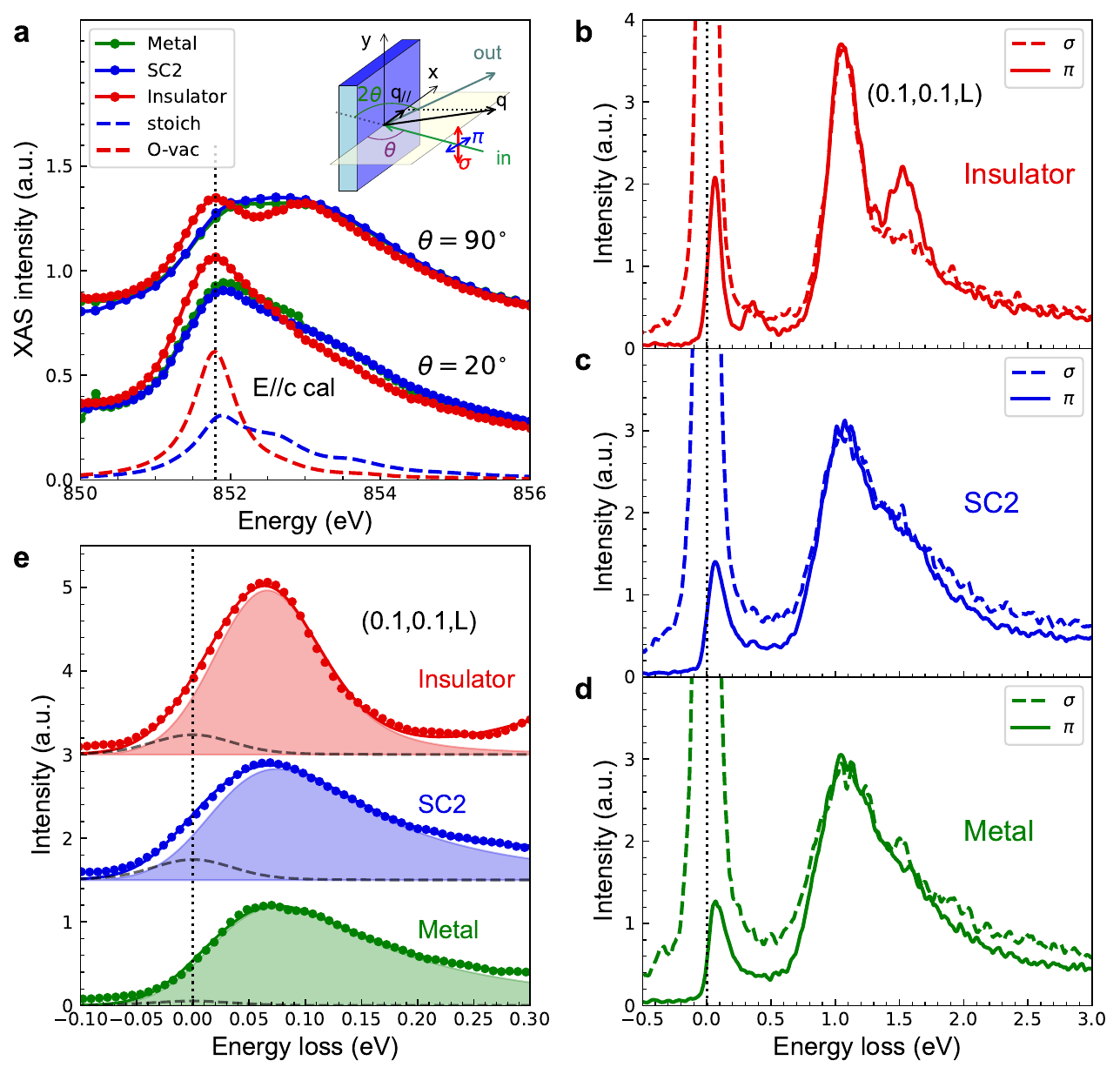}
    \caption{\textbf{Ni $L_3$-edge XAS and RIXS spectra.} \textbf{a}, Ni $L_3$-edge XAS measured at normal incidence and $20^\circ$ grazing incidence with $\pi$ polarization, predominantly probing in-plane and out-of-plane orbitals, respectively. The inset illustrates the measurement geometry. Dashed lines show calculated $E \parallel c$ spectra based on two-cluster models for stoichiometric and inner-apical-oxygen-deficient configurations. \textbf{b--d}, Polarization-dependent RIXS spectra of insulating, metallic, and superconducting samples measured at a fixed scattering angle $2\theta = 90^\circ$. \textbf{e}, Shaded regions highlight low-energy magnetic excitations in the $\pi$-polarized RIXS spectra. The dashed line denotes the fitted elastic peak, strongly suppressed under $\pi$-polarization at the $2\theta = 90^\circ$ condition.
}
    \label{fig2}
\end{figure}

\begin{figure}[t]
    \centering
    \includegraphics [width = \columnwidth]
    {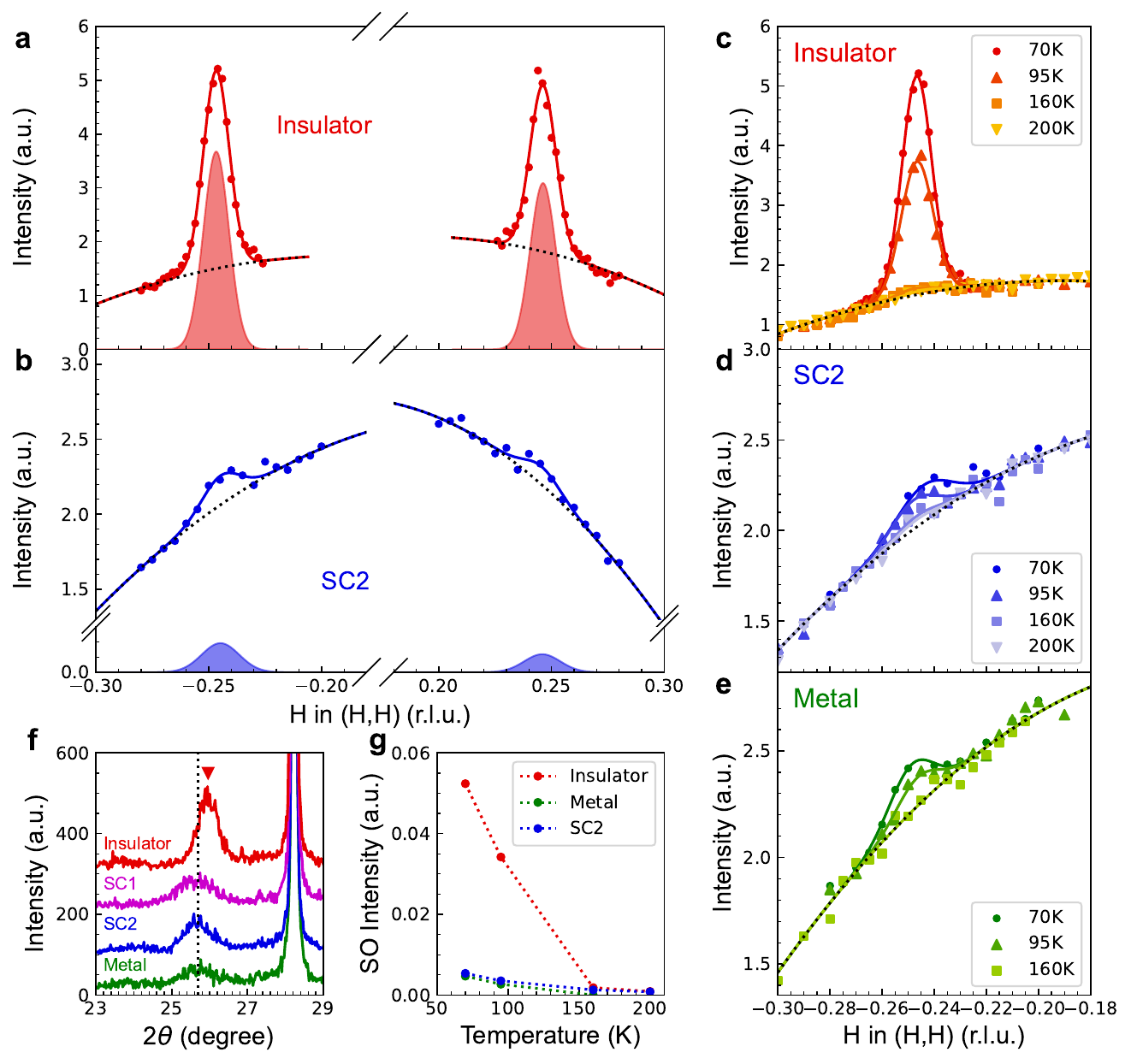}
    \caption{\textbf{Evolution of spin density wave.} \textbf{a,b}, Integrated elastic intensity along the $(H,H)$ direction for positive and negative momentum transfers under $\pi$ polarization. Red and blue curves are Gaussian fits on a quadratic background (gray dashed lines). A pronounced, symmetric SDW peak is observed in the insulating sample, whereas it is strongly suppressed in the superconducting sample. \textbf{c--e}, Temperature dependence of the integrated elastic intensity for the insulating, metallic, and superconducting samples, respectively. \textbf{f}, X-ray diffraction profiles along the $c$ axis, showing a reduced lattice parameter in the insulating sample, indicative of partial strain relaxation. \textbf{g}, Temperature dependence of the SDW/SDF peak intensity extracted from the Gaussian fits.
}
    \label{fig3}
\end{figure}

\begin{figure}[t]
    \centering
    \includegraphics [width = \columnwidth]
    {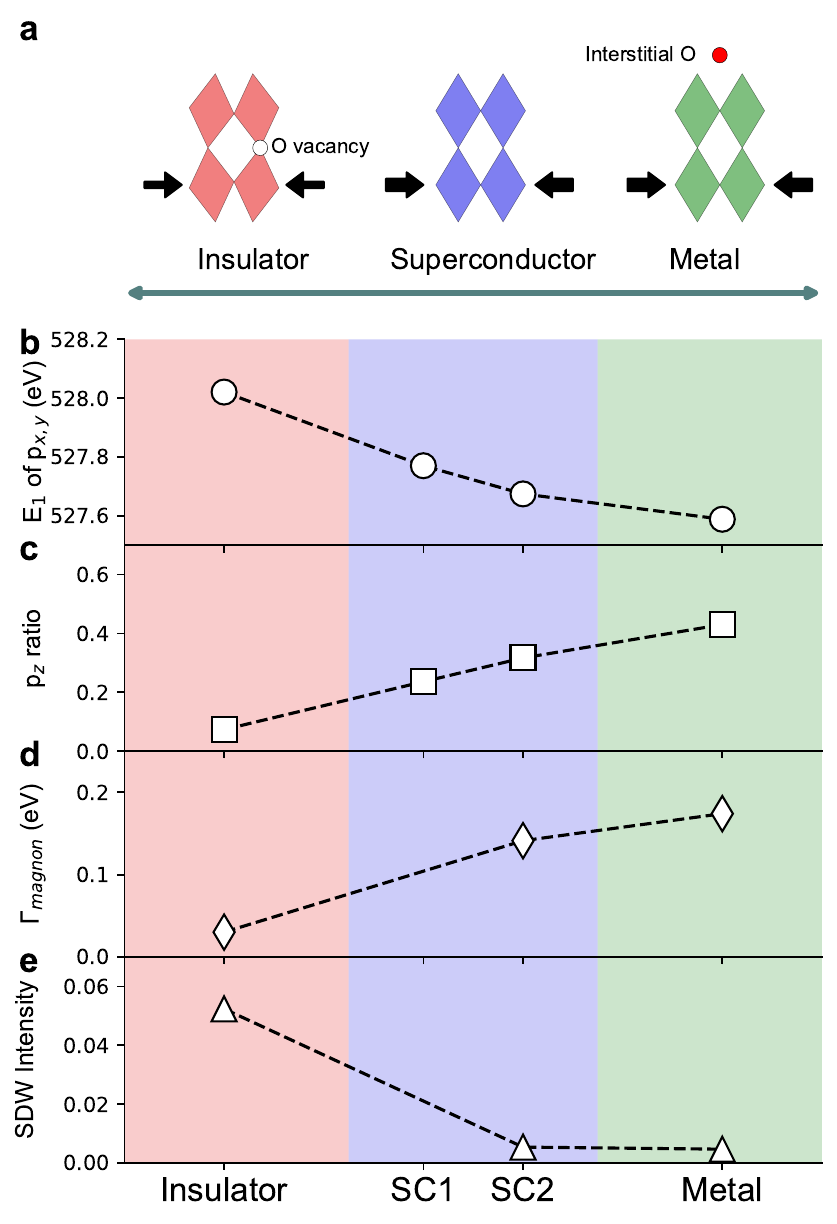}
    \caption{\textbf{Evolution from insulator to superconductor to metal.} \textbf{a}, Schematic illustration of the three phases. The insulating sample contains inner apical oxygen vacancies, exhibits partial strain relaxation, and resembles the ambient-pressure bulk structure. The superconducting and metallic samples are under higher epitaxial strain, with the metallic phase further containing interstitial oxygen. \textbf{b}, Energy of the leading peak in the in-plane O $K$-edge XAS, reflecting the Fermi level position and indicating increasing hole doping from insulator to superconductor to metal. \textbf{c}, Intensity ratio of the leading peak to the total pre-edge in the out-of-plane O $K$-edge XAS, representing the density of states near $E_F$ associated with the Ni $d_{z^2}$ orbital. The increase signals progressive $d_{z^2}$ metallization. \textbf{d}, Damping parameter $\Gamma$ of magnetic excitations extracted from Ni $L_3$-edge RIXS, showing enhanced damping with increasing itinerancy. \textbf{e}, Integrated SDW peak intensity, demonstrating the suppression of magnetic order upon entering the superconducting and metallic regimes.
}
    \label{fig4}
\end{figure}

~\\
\leftline{\textbf{Ni L$_3$-edge XAS and RIXS spectra}}

To probe the evolution of orbital and spin degrees of freedom, we measured Ni $L_3$-edge XAS and RIXS spectra. Figure~2a shows $\pi$-polarized XAS at normal and grazing incidence, which predominantly probe in-plane and out-of-plane orbitals, respectively. The insulating sample exhibits two characteristic peaks, consistent with prior reports and assigned to predominantly $d^8$ and $d^8\underline{L}$ ($\underline{L}$, ligand hole) states at lower and higher energies~\cite{SDW_RIXS_LNO327_NC_2024,PhysRevB_111_014515}. In the superconducting and metallic samples, the in-plane $d^8$ peak broadens and shifts to higher energy, indicating an increased Ni valence and consistent with the hole-doping trend inferred from O $K$-edge XAS. In contrast, the insulating sample shows a reduced high-energy $d^8\underline{L}$ spectral weight for $E \parallel c$, indicative of suppressed Ni--O hybridization along the $c$ axis. This behavior is well captured by calculations comparing stoichiometric and inner apical oxygen-deficient structures, supporting the presence of inner apical oxygen vacancies in the insulating sample. Notably, the Ni $L_3$-edge XAS remains nearly unchanged between the superconducting and metallic samples, in contrast to the pronounced evolution observed at the O $K$-edge. This dichotomy mirrors the doping evolution in $\mathrm{La_{2-x}Sr_xCuO_{4+\delta}}$~\cite{LSCO_XAS_doping_PRL1991_CTC,LSCO_XAS_pol_PRL1992_CTC}, and supports the classification of La$_3$Ni$_2$O$_7$ as a charge-transfer system in which doped holes primarily reside on oxygen ligands.

RIXS spectra further reveal a marked evolution of orbital excitations (Figs.~2b-d). In the insulating sample, three well-defined peaks appear at  $\sim$0.4~eV, $\sim$1~eV and $\sim$1.5~eV consistent with previous studies on single crystals and assigned to orbital excitations ~\cite{SDW_RIXS_LNO327_NC_2024,RIXS_magnon_arxiv_xingye}. Additonally, the $\sim$0.4~eV and $\sim$1.5~eV peaks exhibit strong polarization dependence. Upon entering the superconducting and metallic regimes, these excitations broaden substantially: the 0.4~eV mode is strongly suppressed, and the 1.5~eV feature loses its polarization dependence, indicating enhanced itinerancy and reduced orbital selectivity. We also identify pronounced magnetic excitations at  $\sim$70~meV at $\mathbf{q}$= (0.1,0.1, L) rlu, as shown in Fig.~2e. The magnon energy remains nearly unchanged upon the evolution from the insulating to superconducting and metallic samples. In contrast, the width increases dramatically, signaling a crossover from well-defined magnons to strongly damped spin excitations due to the increase of itinerant carriers, consistent with the progressive metallization of the system.

~\\
\leftline{\textbf{Evolution of spin density wave}}

To elucidate the relationship between spin density wave (SDW) order and superconductivity, we extract the momentum dependence of the quasi-elastic peak intensity along the $(H,H)$ direction from the RIXS spectra. A pronounced SDW peak is observed near $(0.25, 0.25)$ r.l.u. in the insulating sample (Fig.~3a), similar to previous reports ~\cite{SDW_RIXS_LNO327_NC_2024,SDW_REXS_LNO327_comphys2025,SDW_REXS_LNO327_NC2025}. Intriguingly, the SDW peak is strongly suppressed in the superconducting samples, yielding weak spin density fluctuations (SDF) (Fig.~3b). Upon warming, both SDW and SDF peak intensity decrease with increasing temperature and vanish above $\sim$160 K  (Fig.~3c-g).  Notably, the SDW wavevector in the insulating sample is determined as $0.246$ r.l.u. when referenced to the substrate lattice parameters ($a=b=3.756$~\AA), but corresponds to $0.250$ r.l.u. when expressed using the bulk lattice constant ($a \approx 3.87$~\AA). This indicates partial strain relaxation in the insulating sample relative to the superconducting and metallic samples, which is supported by X-ray diffraction measurements showing a shorter $c$-axis in the insulating sample (Fig.~3f).

~\\
\leftline{\large{\textbf{Discussion}}}

The consistency between O $K$-edge XAS and our calculations provides compelling evidence that the insulating sample hosts inner apical oxygen vacancies. This interpretation is further supported by the close resemblance of both the RIXS spectra and SDW signal to those reported in bulk, ambient-pressure La$_3$Ni$_2$O$_7$ single crystals~\cite{SDW_RIXS_LNO327_NC_2024,SDW_REXS_LNO327_comphys2025,SDW_REXS_LNO327_NC2025}, where such vacancies have been directly identified by EELS~\cite{Oxygen_EELS_LNO327_Nat2024}. The removal of inner apical oxygen disrupts the interlayer Ni--O--Ni network, suppressing $d_{z^2}$--$p_z$--$d_{z^2}$ hybridization and effectively breaking interlayer valence bonding. As a result, $d_{z^2}$ electrons become localized, giving rise to magnetic moments and Kondo-like scattering, thereby stabilizing the insulating state, consistent with theoretical proposals~\cite{LNO327_theory_interlayer_PRB2023_ZFC,LNO327_Theory_pair_PRL2023_YF,theory_SC_vanc_LNO327_CP_yang}. These findings reveal oxygen vacancy disorder as a key driver of enhanced electronic correlations and magnetic localization.

As the system evolves from insulator to superconductor and eventually to a metallic state, we observe a continuous increase in hole doping (Fig.~4b), accompanied by enhanced spectral weight near $E_F$ along the $c$ axis (Fig.~4c). The latter is quantified by the intensity ratio between the leading peak and the pre-edge features in the out-of-plane O $K$-edge XAS, reflecting progressive metallization of the $d_{z^2}$ orbital. This evolution is not driven solely by carrier concentration, but by the emergence of interlayer electronic coherence via $d_{z^2}$--$p_z$--$d_{z^2}$ hybridization. Consistently, hole doping shifts $E_F$ and is accompanied by the broadening and suppression of the $\sim$0.4~eV feature, previously attributed to interorbital excitations between $d_{z^2}$ and $d_{x^2-y^2}$ states~\cite{SDW_RIXS_LNO327_NC_2024}. Moreover, the reduced polarization dependence of orbital excitations in the superconducting and metallic samples indicates enhanced interlayer hybridization and a crossover toward more three-dimensional electronic character. Our results thus reveal distinct yet cooperative roles of epitaxial strain and oxygen stoichiometry. Epitaxial strain straightens the Ni--O--Ni bond angle toward $180^\circ$, enhancing interlayer hopping, but is not sufficient to induce superconductivity. Instead, optimal oxygen stoichiometry is required to activate coherent $d_{z^2}$--$p_z$--$d_{z^2}$ hybridization, thereby tuning both carrier density and correlation strength. The establishment of this coherent interlayer electronic network is the key condition for superconductivity.

Concomitantly, the magnetic excitation spectrum exhibits increased damping upon entering the superconducting and metallic states (Fig.~4d), indicative of enhanced electronic itinerancy. Notably, the characteristic magnetic energy scale remains largely unchanged, indicating that the underlying spin exchange interaction is robust across the phase diagram. This behavior parallels that observed in cuprates, where magnetic excitations persist deep into the overdoped regime~\cite{magnon_HTc_NP2011_BK,magnon_LSCO_NM2013_MPM,magnon_odcuprate_PRB2013_BK,magnon_bi_PRB2018_YYP}. In contrast, the SDW order is strongly suppressed in the superconducting and metallic samples (Fig.~4e), consistent with $\mu$SR and optical studies under pressure~\cite{DW_press_ultrafast_NC2024,SDW_muSR_LNO327_PRL_2024,SDW_pressure_muSR_LNO327_NP2025,SDW_SC_pressure_untrafast_NC2025,SDW_pressure_muSR_LNO327_PRR2025}. These observations suggest that SDW constitutes a competing ground state in the insulating regime, while its fluctuations persist into the superconducting phase and may contribute to pairing.

Further insight into the doping evolution is provided by the metallic sample, where epitaxial strain remains essentially unchanged compared with the superconducting sample (Fig.~3f), yet additional hole carriers are introduced. Our calculations attribute this overdoping to interstitial oxygen (Fig.~1f,g), consistent with EELS observations in metallic La$_3$Ni$_2$O$_7$~\cite{oxygen_EELS_LPNO2127_NM2025}. Despite the increased carrier density, the similarity of the XAS and RIXS spectra with those of superconducting sample indicates that interstitial oxygen does not qualitatively modify the orbital character or magnetic excitations, but instead weakens electronic correlations by driving the system into an overdoped regime and suppressing superconductivity. Together with the vacancy-induced localization in the insulating state, this establishes a unified picture in which oxygen stoichiometry simultaneously controls carrier density and correlation strength. This scenario is consistent with recent theories proposing a dual instability of superconductivity from oxygen defects, where vacancies promote localization and magnetism, while interstitials yield a coherent but weakly correlated metallic state unfavorable for superconductivity~\cite{theory_Odefect_Arxiv2025_WHQ,theory_Odefect_Arxiv2026_YF}. 

In summary, our results reveal that the evolution from insulating to superconducting to metallic states in La$_3$Ni$_2$O$_7$ is predominantly manifested in the out-of-plane electronic structure, through the activation of coherent interlayer $d_{z^2}$--$p_z$--$d_{z^2}$ bonding within the bilayer, while the in-plane $d_{x^2-y^2}$ states evolve gradually. Oxygen stoichiometry and strain cooperatively tune hole density and interlayer coupling, thereby controlling orbital-selective electronic states and suppressing competing orders. As a result, superconductivity emerges only within a narrow window of oxygen stoichiometry, even under otherwise favorable structural conditions, where  static SDW order is suppressed while robust spin exchange interactions persist. These findings establish oxygen as a dual tuning parameter governing both electronic coherence and correlation strength, and provide a multiorbital framework for superconductivity in bilayer nickelates. Our results provide insight into materials-design strategies that decouple carrier doping from correlation tuning--for example, by compensating interstitial-oxygen-induced hole doping with electron substitution at the La site--to expand the accessible synthesis window and stabilize superconductivity.

~\\
\leftline{\large{\textbf{Methods}}}\\

\small{
\noindent{\textbf{Thin Film Fabrication}}\\
Epitaxial LaPr$_2$Ni$_2$O$_7$ thin films were grown by pulsed laser deposition (PLD) using a KrF excimer laser ($\lambda = 248$ nm) on (001)-oriented SrLaAlO$_4$ single-crystal substrates (Hefei Kejing Co., Ltd.). A polycrystalline LaPr$_2$Ni$_2$O$_7$ target was prepared by conventional solid-state reaction. During deposition, the substrate temperature was maintained at 780~$^\circ$C, with a laser repetition rate of 4 Hz and a fluence of 1.8 J\,cm$^{-2}$. Growth was performed in a 10\% O$_3$/90\% O$_2$ atmosphere at a total pressure of 0.41 mbar. After deposition, the films were cooled to room temperature at a rate of 50--100~$^\circ$C\,min$^{-1}$ under the same atmosphere. For thicker films, an additional in situ annealing step was carried out at 400~$^\circ$C for 10--20 min prior to cooling. The film thickness was controlled by laser pulse counting and calibrated using X-ray reflectivity (XRR). Finally, an amorphous PrBa$_2$Cu$_3$O$_7$ capping layer ($\sim$1 nm) was deposited ex situ at room temperature under identical atmospheric conditions, using a reduced laser fluence of 1.0 J\,cm$^{-2}$.

~\\
\textbf{XRD and Electrical Transport Measurements}\\
The crystallographic quality of the films was assessed by high-resolution X-ray diffraction (XRD) $\theta$--$2\theta$ scans using a 3 kW Rigaku SmartLab diffractometer with a monochromatic Cu K$_\alpha$ source. For transport measurements, 50 nm-thick Ag electrodes were deposited on the film surface by electron-beam evaporation, and electrical contacts were established via ultrasonic Al wire bonding. The transport properties were measured in a standard four-probe configuration using a commercial Physical Property Measurement System (PPMS, Quantum Design).

~\\
\textbf{XAS and RIXS measurements}\\
X-ray absorption spectroscopy (XAS) and resonant inelastic X-ray scattering (RIXS) experiments were carried out at the beamline 41A of the Taiwan Photon Source (TPS), National Synchrotron Radiation Research Center, Taiwan~\cite{TPS41A}. The in-plane lattice parameters were defined by the substrate ($a=b=3.756$~\AA), and the measurement geometry is illustrated in Fig.~2a. For the SC2, metallic, and insulating samples, measurements were performed in the $(H,H,L)$ scattering plane, whereas for SC1 the $(H,0,L)$ plane was used. O $K$-edge XAS spectra were collected in total fluorescence yield (TFY), and Ni $L_3$-edge XAS in total electron yield (TEY).

RIXS measurements were conducted at the Ni $L_3$ resonance energy under grazing-incidence $\pi$ polarization (Fig.~2a), with an overall energy resolution of $\sim$85 meV. For the spectra shown in Fig.~2, the scattering angle $2\theta$ was fixed at $90^\circ$, while for SDW measurements (Fig.~3) it was set to $150^\circ$ under $\pi$ polarization. Unless otherwise specified, all XAS and RIXS data were acquired at 70 K for SC2, metallic, and insulating samples, and at 30 K for SC1, except for the temperature-dependent measurements shown in Fig.~3c--e.

~\\
\textbf{XAS Data Analysis}\\
The O $K$-edge XAS spectra shown in Fig.~1d,e were obtained by decomposing measurements acquired at different polarization and incidence angles; details are provided in the Supplementary Information. To enable comparison across samples, the spectra were normalized to the main peak intensity. The spectral analysis was performed using a Lorentzian function centered at 531.8 eV to describe the main peak background, together with two Gaussian components to fit the pre-edge features.

For the Ni $L_3$-edge XAS, spectra at normal and grazing incidence were first fitted using a linear background and three Lorentzian peaks, corresponding to the La $M_4$ edge and two Ni $L_3$ components. After subtracting the La $M_4$ contribution and background, the Ni $L_3$ spectra were normalized and are presented in Fig.~2a. Further details of the fitting procedure are provided in the Supplementary Information.

~\\
\textbf{RIXS Data Analysis}\\
The RIXS spectra in Fig.~2 were normalized to the integrated intensity between 0.8 and 1.2 eV, dominated by the $\sim$1 eV $dd$ excitation. For the fitting in Fig.~2e, the spectra were modeled using a linear function for the fluorescence background, a resolution-limited peak centered at zero energy loss for the elastic line, a Lorentzian component for the $\sim$0.4 eV excitation, and an anti-Lorentzian function convoluted with the instrumental resolution to describe the magnetic excitation. The energy-resolution function was determined from measurements at the specular reflection condition. For Fig.~3, the spectra were first normalized to the integrated intensity between 0.5 and 2 eV. The quasi-elastic intensity was then obtained by integrating the spectral weight within $\pm$0.1 eV around zero energy loss. 
~\\
\textbf{XAS calculations}\\
O $K$-edge XAS were simulated for Pr$_3$Ni$_2$O$_7$ and its variants using the excited electron and core-hole approach as implemented in the Vienna ab initio Simulation Package (VASP)~\cite{VASP1,VASP2}, with the Perdew-Burke-Ernzerhof (PBE) functional~\cite{PBE} and a Hubbard $U = 4$~eV on the Ni~$3d$ orbitals~\cite{Dudarev1998}. The in-plane lattice constant was fixed to that of the SrLaAlO$_4$ substrate, while the $c$-axis parameter and internal atomic positions were fully relaxed. Three oxygen stoichiometries were considered, including stoichiometric Pr$_3$Ni$_2$O$_7$, oxygen-vacant Pr$_3$Ni$_2$O$_{6.75}$, and oxygen-rich Pr$_3$Ni$_2$O$_{7.25}$, with vacancies at the inner-apical site and interstitials in the inter-bilayer rock-salt layer, as indicated by EELS experiments~\cite{Oxygen_EELS_LNO327_Nat2024,oxygen_EELS_LPNO2127_NM2025} and earlier DFT calculations~\cite{theory_Odefect_Arxiv2025_WHQ}. Spectra were computed in a $2\times2\times1$ supercell. Further details are given in Supplementary Note~5.

Ni $L$-edge XAS were calculated using ligand-field multiplet theory with a two-cluster model comprising two NiO$_6$ octahedra coupled through the shared inner-apical oxygen, as previously employed in Ref.~\cite{SDW_RIXS_LNO327_NC_2024}. The calculation fully correlates the Ni $3d$ and surrounding O $2p$ shells, with a
charge-transfer energy $\Delta = 0.5$~eV, and Slater integrals scaled to 80\% of their atomic values. For the stoichiometric case, the nominal Ni occupation is $d^{7.5}$; for the oxygen-vacant variant, the inner apical oxygen was removed, decoupling inter-cluster hybridization and yielding a nominal occupation of $d^{8.5}$. Calculations were performed using Quanty~\cite{Quanty}. Detailed model construction and parameter sets are given in Supplementary Note~6.
}

~\\
\leftline{\large{\textbf{Data availability}}}
\small{
The data used to support the findings of this work are available from the corresponding
author upon request. Source data are provided with this paper.
}

~\\
\leftline{\large{\textbf{Acknowledgements}}}
\small{
We thank George Sawatzky, Fan Yang, Fa Wang, Ilya M. Eremin for their helpful discussions. Y.Y.P. is grateful for financial support from the National Natural Science Foundation of China (Grants No.12374143), the Ministry of Science and Technology of China (Grants No. 2021YFA1401903 and No. 2024YFA1408702), and Beijing Natural Science Foundation (Grant No. JQ24001).  Y.L. acknowledges financial support from the Ministry of Science and Technology of China (Grant No. 2022YFA1403000), the National Natural Science Foundation of China (Grant No. 12274207), and the Basic Research Program of Jiangsu (No. BK20253009). Y.X. acknowledges support from the NSFC (Grant No. 12325402).  M.Z. acknowledges support from the NSFC (Grant No. 12504226).
}

~\\
\leftline{\large{\textbf{Author contributions}}}
\small{
S.Z. and Y.P. proposed and designed the research. S.Z., Z.T., H.H., K.L., D.H. and Y.P. carried out the RIXS and XAS experiments. M.Z., J.F. and Y.X. provided the samples and M.Z. carried out the XRD and resistance experiments. Q.L., J.L. and Y.L. performed theoretical calculations. S.Z., Y.L. and Y.P. prepared the manuscript. All authors have read and approved the final version of the manuscript.
}

~\\
\leftline{\large{\textbf{Competing Interests}}}
\small{
The authors declare no competing interests.
}

\end{document}